
\input harvmac.tex
%
%
\def\havefigures{n}


\def\figures{y}
\ifx\havefigures\figures
	\input epsf.tex
	\def\topfigure#1#2#3{\topinsert {\centerline{\epsffile{#1}}}
	\smallskip\tenpoint\baselineskip12pt \noindent
	{\bf Fig.~{#2}.} {\rm \ \  {#3}} \endinsert}
\else
	\def\topfigure#1#2#3{}
\fi


\def\preprint#1{\nopagenumbers\abstractfont\pageno=0
\hsize=\hstitle\rightline{#1}}

\def\date#1{\rightline{#1}}

\def\title#1#2{
   \vskip 1in
   \centerline{\titlefont #1}
   \vskip 0.2in
   \centerline{\titlefont #2}
   \vskip 0.5in plus 0.1in}

\def\author#1#2#3{\smallskip\centerline{{\bf #1}\footnote{#2}{#3}}\smallskip}
\def\address#1{\centerline{#1}}

\def\abstract{\centerline{\bf Abstract}\smallskip}
\def\finishtitlepage{\Date{\ }}


\def\monthintext{\ifcase\month\or January\or February\or
   March\or April\or May\or June\or July\or August\or
   September\or October\or November\or December\fi}


\let\ulabelfoot=\footnote


\def\undersection#1{\par
   \ifnum\the\lastpenalty=30000\else \penalty-100\medskip \fi
   \noindent\undertext{#1}\enspace \vadjust{\penalty5000}}
\def\undertext#1{\vtop{\hbox{#1}\kern 1pt \hrule}}
\def\subsection#1{\undersection{#1} \medskip}


\def\ack{\ifnum\the\lastpenalty=30000\else \penalty-100\smallskip \fi
   \noindent\undertext{Acknowledgements:}\enspace \vadjust{\penalty5000}}


\def\en{\eqalign}
\def\del{\partial}


\def\en{\eqalign}

\def\cE{ {\cal E} }

\def\cO{ {\cal O} }

\def\cH{ {\cal H} }

\def\cP{ {\cal P} }
\def\state#1{ {| #1\rangle} }
\def\lstate#1{ {\langle #1 |} }

\nref\mrobertson{G.~McCartor and  D.~Robertson, Z.~Phys.~C53 (1992)679.}
\nref\blone{M.~Burkardt and A.~Langnau,  Phys.~Rev.~D44 (1991)1187.}
\nref\bltwo{M.~Burkardt and A.~Langnau,  Phys.~Rev.~D44 (1991)3857.}
\nref\pinsky{D.~Mustaki, S.~Pinsky, J.~Shigemitsu, \hfill\break
K.~Wilson,~Phys.~Rev.~D43~(1991)3411.}
\nref\thorn{C.~Thorn, Phys.~Rev.~D20 (1979)1934.}
\nref\kent{K. Hornbostel, Phys.~Rev.~D45 (1992)3781.}
\nref\llty{F.~Lenz, S.~Levit, M.~Thies, and K.~Yazaki,
Ann.~Phys.~208~(1990)1.}
\nref\franke{E.~Prokvatilov and V.~Franke, Yad.~Fiz.~49 (1989)1109.}
\nref\changma{S. Chang and S. Ma, Phys.~Rev.~180 (1969)1506.}
\nref\ksop{J.~Kogut and D.~Soper, Phys.~Rev.~ D1 (1970)2901.}
\nref\bks{J.D.~Bjorken,J.~Kogut, and D.~Soper, Phys.~Rev.~ D3 (1971)1382.}
\nref\cry{S.~Chang, R.~Root, and T.~Yan, Phys.~Rev.~D7 (1973)1133,1147.}
\nref\mandel{S. Mandelstam, Nucl.~Phys.~B213 (1983)149.}
\nref\leibbrandt{G. Leibbrandt, Phys.~Rev.~D29 (1984)1699.}
\nref\colemansg{S.~Coleman, Phys.~Rev.~D11 (1975)2088.}
\nref\samuel{S.~Samuel, Phys.~Rev.~D18 (1978)1916.}
\nref\mrg{P.~Minnhagen, A.~Rosengren, and G.~Grinstein,
Phys.~Rev.~B18(1978)1356.}
\nref\agg{D.~Amit, Y.~Goldschmidt, and G.~Grinstein, J.~Phys.~A13 (1980)585.}
\nref\stf{E.~Sklyanin, L.~Takhtadzhyan, and L.~Faddeev, Teor.~Mat.~Fiz.
40 (1979)194.}
\nref\miransky{V.A.~Miransky, Nuovo~Cimento~90A~(1985)149.}
\nref\gellow{M.~Gell-Mann, and F.~Low, Phys.~Rev.~84(1951)350.}
\nref\bpa{S.~Brodsky and H.C.~Pauli, Phys.~Rev.~D32 (1985)1993.}
\nref\hv{A.~Harindranath and J. Vary, Phys.~Rev.~D37 (1988)3010.}

\preprint{UFIFT-HEP-17}
\date{May 1992}
\title	{The sine-Gordon model and the small $k^+$ region}
	{of light-cone perturbation
	theory\ulabelfoot{{\titlefont$^\dagger$}}
	{Supported in part by the U.S.  Department of Energy,
	under grant DE-FG05-86ER-40272} }
\author	{Paul A. {Grif}fin} {$^{\dagger\dagger}$}
	{Internet: pgriffin@ufhepa.phys.ufl.edu}
\address{Department of Physics, University of Florida}
\address{Gainesville, FL 32611}
\abstract
The non-perturbative ultraviolet divergence of the sine-Gordon model is used
to study the $k^+ = 0$ region of light-cone perturbation
theory.
The light-cone vacuum is shown to be unstable at the non-perturbative
$\beta^2 = 8\pi$ critical point by a
light-cone version of Coleman's variational method.
Vacuum bubbles, which are $k^+=0$ diagrams in light-cone
field theory and are individually finite and non-vanishing for all $\beta$,
conspire to generate ultraviolet divergences of the light-cone energy
density.
The $k^+ = 0$ region of
momentum also contributes to connected Green's functions; the connected
two point function will not diverge, as it should, at the critical point
unless diagrams which contribute only at $k^+ = 0$ are properly included.
This analysis shows in a simple way how the $k^+ =0$ region cannot
be ignored even for connected diagrams.  This phenomenon is expected to
occur in higher dimensional gauge theories starting at two loop order in
light-cone perturbation theory.

\finishtitlepage

\newsec{Introduction}

Recently, much attention has been given to the issue of regulating
the non-covariant divergences which occur in canonical light-cone field
theory\mrobertson\blone\bltwo\pinsky.  In the naive application of this
perturbation theory, the $k^+ = 0$ region of light-cone Feynman diagrams
is regulated by applying a cutoff to $k^+ \sim \epsilon$ to momentum
integrals.  It is then assumed that for physical (gauge invariant)
processes the $\epsilon \rightarrow 0$ limit can be taken at the
end of the full calculation, i.e.~the $k^+ = 0$ region does not
contribute to physical processes.  This assumption is analogous to the
(in this case correct) expectation that infared (IR) divergences cancel in
inclusive processes for theories with massless particles.  In fact,
a straight cutoff of the $k^+ = 0$ region does give the right
renormalization structure at one loop for QED\pinsky\ and the right beta
function $\beta (g)$ for QCD\thorn.
Furthermore, it is standard lore that the light-cone vacuum is trivial;
i.e.~that the interacting vacuum is the free-field vacuum.  In
light-cone perturbation theory this implies that all
bubble diagrams, which have support only in the $k^+ = 0$ region, actually
vanish; certainly with a sharp $k^+$ cutoff they can not contribute.
However, that the light-cone vacuum actually must be non-trivial has been
discussed
recently in refs.~\kent\llty\franke, and in fact was actually realized many
years ago\changma.

In this letter, the sine-Gordon model in two dimensions will be used to
explore these questions.  We will see that vacuum bubbles do not
vanish, and (more importantly) that the $k^+ = 0$ region of light-cone
perturbation theory contributes to connected Green's functions at
second order in the light-cone Hamiltonian perturbation theory;
these facts are intimately related.  For the more physical gauge theories in
four dimensions, it will be clear that the $k^+ = 0$ region can contribute to
connected Green's functions at two-loop order and beyond.

The analysis of this paper is in the framework of canonical quantization
on the light-cone (null plane) $x^+ = (t+x)/ \sqrt 2 =0$, and Dyson's
(old fashioned) Hamiltonian perturbation theory for the
light-cone\ksop\bks\cry.  While the issue addressed here is not the same
as the $k^+ = 0$ divergence in the covariant Minkowski-space
Feynman perturbation theory of gauge theories in the light-cone gauge,
which is treated with the Mandelstam-Leibbrandt
prescription\mandel\leibbrandt, they are probably deeply related.

Note that no new properties will discovered about the sine-Gordon model per se,
and only sine-Gordon perturbation theory\colemansg\samuel\mrg\agg\ will
be discussed.  (The exact sine-Gordon solution via the inverse
scattering method in given in ref.~\stf.)  The reason for using this
model is that the $k^+ = 0$ issue appears to lowest non-vanishing
order in sine-Gordon perturbation theory, and is easy
to calculate because the sine-Gordon model is ultraviolet (UV) finite diagram
by
diagram\colemansg.
In addition, we will use existence of the phase transition
at $\beta^2 = 8\pi$ as a check on the validity of light-cone perturbation
theory.

\newsec{The instability at $\beta^2 = 8\pi$}
To warm up, let us calculate the critical point of the sine-Gordon model
using Coleman's variation method\colemansg\ and light-cone quantization.
This will lead to insight on the nature (IR verses ultraviolet UV ) of the $k^+
= 0$ singularity.
In particular, we want to show that for $\beta^2 \geq 8\pi$, the Hamiltonian
density $\cH$ is unbounded from below.

For canonical light-cone quantization on the null-plane, the momentum operator
is $P^+ =\int dx^- (\cH + \cP )/\sqrt 2 $ and the Hamiltonian operator is
$P^- =
\int dx^- (\cH - \cP)/\sqrt 2$.  For the sine-Gordon model, they are
given by
\eqn\engmom{
\en{
P^+ =& \int dx^- \del_- \phi \del_- \phi \ , \cr
P^- =& \int dx^- {\alpha_0\over \beta^2}(1- \cos \beta\phi )
 \ , \cr }
}
where $\alpha_0$ and $\beta$ are taken to be positive\colemansg.
If we expand about the configuration of minimum $P^-$, ($\phi = 0$),
then $1- \cos \beta\phi$ is an even power series in  $\phi$; the
quadratic term has coefficient $\half \alpha_0 =\half m^2 $, where
$m$ is mass associated with perturbation theory about $\phi = 0$.
The canonical boson field has mode expansion at $x^+ = 0$
\eqn\expand{
\phi = {1\over \sqrt {4\pi}} \int^\infty_0 {dk^+\over k^+}
\left[ a( k^+ )e^{-ik^+ x^-} + a^\dagger( k^+ ) e^{+ik^+ x^-}\, \right]\ ,
}
where the $1/k^+$ term in the integrand of eqn.~\expand\ comes from the
covariant measure for free particles.
The canonical commutation relations are
\eqn\ccr{
[ a (k^+ ), a^\dagger (k^{\prime +} )] = k^+ \delta ( k^+ - k^{\prime +} ) \ .
}
In terms of the creation $a^\dagger$ and annihilation $a$ operators, the
light-cone momentum density operator is
\eqn\peeplus{
N_m[ \del_-\phi \del_- \phi ] + {1\over 4\pi} \int^\infty_0 dk^+ k^+ \ ,
}
where normal ordering $N_m [\ ]$ is with respect to the free-field
vacuum with mass $m$.
In order to perform a variational calculation of the
Hamiltonian density with respect a new vacuum state $\state \mu$ that
corresponds to different free-field mass $\mu$, it necessary to calculate the
divergent part of eqn.~\peeplus\ with respect to a UV momentum
cutoff\colemansg.

To this end, introduce
``IR" and ``UV" cutoffs $\delta^+$ and $\Lambda^+$,
\eqn\cutoffs{
\int^\infty_0 dk^+ \rightarrow \int^{\Lambda^+}_{\delta^+} dk^+  \ .
}
These cutoffs are actually related by parity.  Consider the solution
to the free-field equation of motion,
\eqn\freebie{
\phi = \int^{\Lambda^+}_{\delta^+} {dk^+ \over k^+}\left[ b( k^+ )
e^{+ik^+ x^-} e^{+ m^2 x^+ / 2k^+} + \hbox{c.c.}  \right] \ .
}
Under parity $x \rightarrow -x$, the modes $b$ transform as
$b( k^+ ) \rightarrow b( m^2 / 2 k^+ ) $,
and the cutoffs transform as
$\Lambda^+ \rightarrow  m^2/2\delta^+ $ and
$\delta^+ \rightarrow  m^2/2\Lambda^+ $.
Hence to maintain parity with our regularization of the divergent
part of $P^+$, let
\eqn\cutconnect{
\delta^+ = { m^2 \over 2 \Lambda^+ } \ .
}
This relation introduces the mass $m$ into the divergent part of
$P^+$.  The regulated free-field light-cone vacuum is therefore sensitive to
the
free-field mass.
And note that the $k^+ = 0 $ region is clearly not just IR, since
parity interchanges the IR region with the UV region of $k^+$.
The UV cutoff $\Lambda^+$ is by Lorentz invariance a function of the
free-field mass and a mass-independent momentum cutoff $\Lambda$,
\eqn\cutcut{
\Lambda^+ = {\Lambda + \sqrt {\Lambda^2 + m^2 }\over \sqrt 2} \ .
}
Using relations \cutconnect\ and \cutcut, the regulated light-cone momentum
is
\eqn\regmom{
P^+ = N_m[ P^+ ] + {1\over 8\pi}\left[ 2\Lambda^2 + m^2 + \cO (m/\Lambda )
\right] \ .
}
To similarly regularize the light-cone energy density, relate the exponentials
in the $\cos {\beta \phi}$ term to their normal ordered forms:
\eqn\exponent{
e^{\pm i\beta \phi} =
e^{\pm i\beta \phi^+ }e^{\pm i\beta \phi^- } e^{\beta^2 [\phi^+ , \phi^- ]/2}=
N_m [e^{\pm i\beta \phi}]
\left( {m^2\over 4\Lambda^2} \right)^{\beta^2\over 8\pi} (1+ \cO (m/\Lambda) )
\ ,
}
where $\phi^+$ and $\phi^-$ contain only raising and lowering operators
respectively.

With eqns.~\regmom\ and \exponent, we can easily reproduce the variational
estimate of Coleman.
In the quantum perturbation theory with respect to the mass $m$
vacuum state, the correct expression for the Hamiltonian density is
\eqn\ham{
\cH = {1\over \sqrt 2} N_m [ \del_- \phi \del_-\phi +
{\alpha\over \beta^2}(1- \cos \beta\phi ) ] \ .
}
Normal ordering the exponential eliminates divergent tadpole terms,
and renormalizes the coupling constant $\alpha_0 \rightarrow \alpha$.
The resulting perturbation theory is UV finite, diagram by diagram\colemansg.
Now consider the variational estimate of the energy density
with respect to the perturbative ground states corresponding to free-field
mass $\mu$.  Using eqns.~\regmom\ and \exponent,
\eqn\estimate{
\lstate \mu \cH \state \mu =
{1\over \sqrt 2} \left[ {1\over 8\pi}(\mu^2 - m^2 )
-{\alpha\over \beta^2} \left( \mu^2 \over m^2 \right)^{\beta^2\over 8\pi}
\right] \ , }
where the cutoff $\Lambda \rightarrow \infty$.
For finite $\alpha$ and $\beta > 8\pi$ the energy density of the $\mu$ vacuum
state is unbounded from below as $\mu$ becomes large, and
the theory as defined has no ground state for these values of $\beta$.

The underlying reason that the sine-Gordon perturbation theory is sick for
$\beta^2 > 8\pi$ is based on the fact that that the anomalous dimension of the
cosine term is $\beta^2 / 4\pi$. For $\beta^2 < 8\pi$ the  anomalous dimension
of the interaction Hamiltonian is less than two and the theory is
super-renormalizable, at $\beta^2 =
8\pi$ the cosine term is a marginal operator and the theory is
renormalizable\agg\miransky,
and for $\beta^2 > 8\pi$ the anomalous dimension is greater than 2,
the dimension of the renormalized coupling $\alpha$ is negative,
and the theory is non-renormalizable.

\newsec{The light-cone vacuum is non-trivial}

We will now try to understand the UV divergence for $\beta^2 \geq 8\pi$
diagrammatically in light-cone perturbation theory.  We proved in the
previous section that the light-cone vacuum is unstable in this regime.
This is possible only if vacuum bubbles are non-vanishing,
and the interacting light-cone vacuum is not the free-field light-cone vacuum.
As we will explicitly see, vacuum bubbles in light-cone field theory
are $k^+ = 0$ diagrams, and the sharp cutoff of the $k^+ = 0$ region used in
variational estimate of the previous section is not a
suitable regulator for these diagrams.

The light-cone Hamiltonian density is broken up into
free and interacting parts,
\eqn\twohams{
\en{
P^-_{\rm free} =& \half m^2 \int dx^- N [ \phi^2 ] \ , \cr
P^-_{\rm int} =& {\alpha\over \beta^2}\int dx^- N[ (1- \cos \beta\phi
+ \half\beta^2 \phi^2 )] \ ,\cr}
}
where $m^2 = \alpha$, and the normal ordering is with respect to $m$.
The Dyson perturbation expansion is defined
for operators in the interaction representation, $\cO (x^- , x^+ )=
e^{iP^-_{\rm free}x^+} \cO (x^- )e^{-iP^-_{\rm free}x^+}$.  In particular,
the free-field Green's function is $G(x,x^\prime ) =\lstate 0
T [ \phi (x^- , x^+ ) \phi (x^{\prime -} , x^{\prime +} )]\state 0$, where
the time ordering $T[\ ]$ is with respect to the light-cone time
$x^+$:
\eqn\green{
\en{
G(x) =& {\Theta (+x^+ ) \over 4\pi} \int_0^{\infty} {dk^+\over k^+}
e^{-i [ k^+ x^- + m^2 x^+ / 2k^+ ] }\cr
+&
{\Theta (- x^+ ) \over 4\pi} \int_0^{\infty} {dk^+ \over k^+}
e^{+i [ k^+ x^- + m^2 x^+ / 2k^+ ] } \ .\cr}
}
The integrals over $k^+$ are well defined because the singularity
at $k^+ = 0$ is cancelled by the $k^+ = \infty$ region.  The result
is just the covariant propagator for both time-like and space-like separations.
The one explicit property of $G(x)$ that is required in the following analysis
is that for $|x |m \ll 1$, where $|x |$ is the
invariant distance, the Green's function is
$G(x) = -\ln [m^2 x^2]/4\pi$.  A standard tool from light-cone perturbation
theory that will be applied is Wick's theorem for $x^+$-ordered exponentials,
\eqn\wick{
 T[ e^{i\int d^2x j(x) \phi(x)} ]
= e^{-\half\int d^2 x d^2 y j(x) j(y) G(x-y)}N[e^{i\int d^2x j(x) \phi(x)} ]\ .
}

The interacting vacuum light-cone energy density $\cE^-$ is given by a
straightforward
rewriting of the Gell-Mann Low equal time formula\gellow\ for the light-cone
case,
\eqn\density{
\cE^- =\lstate {0} \cP^-_{\rm int}(0)\, T\left[ \exp  { \left(
 -i\int^0_{-\infty} dx^+ \cP^-_{\rm int}\right) }  \right]
\, {\state 0}_{\rm conn} \ ,
}
where $\lstate 0 \cdots {\state 0}_{\rm conn}$ is the connected (to
the light-cone Hamiltonian density $\cP^-_{\rm int}(0)$)
free-field vacuum expectation value.  This expression is a perturbation
theory in $\alpha$.  To first order in $\alpha$,
$\cE^-_1= \lstate 0 P^-_{\rm int}(0)\state 0 =0$; but to order $\alpha^2$, we
get the non-vanishing result
\eqn\etwo{
\cE^-_2 = -i{\alpha^2\over \beta^4}\int^0_{-\infty} dx^+
\int^{\infty}_{-\infty} dx^-
\left[ \cosh {(\beta^2 G(x))} - \beta^4 G^2 (x) - 1 \, \right]
\ .
}

\topfigure {sgfig1.eps} {1} {The $x^+$ ordered light-cone diagrams for the
light-cone energy density $\cE^-$, to second order in $\alpha$.}

To recover the light-cone perturbation theory expression for
$\cE^-_2$, expand eqn.~\etwo\ in a power series
in $\beta$, and integrate over coordinates $x^+$ and $x^-$.  Using eqn.~\green,
one easily
finds
\eqn\twolc{
\cE^-_2 = -{\alpha^2\over \beta^4}
\sum_{n=2}^{\infty} {\beta^{2n} \over (2n)!} \int^{\infty}_0
\prod_{p=1}^{n}{dk^+_p \over k^+_p}\  {\delta (\sum_{q=1}^{n} k^+_q )
\over \sum_{r=1}^{n} {m^2\over 2 k^+_r} -i\epsilon } \ .
}
This power series expansion in $\beta$ has the diagrammatic interpretation
shown in figure 1.  It is an infinite sum of two-point connected vacuum
bubbles,
which are non-vanishing only in the $k^+ \rightarrow 0$ limit.  Because
of the ratio $\delta (\sum k^+ ) / \prod k^+$, this limit is ill-defined;
regulating the $k^+ = 0$ region by introducing
a cutoff $k^+ \geq \delta^+$ would miss this contribution altogether.
How one should properly evaluate these integrals is no mystery however, since
the coordinate space representation given by eqn.~\etwo\ is perfectly
well defined.

To see that eqns.~\etwo\ and \twolc\ must be non-vanishing, consider
the non-perturbative UV divergence of the sine-Gordon model.
The divergence occurs when the point separation $|x|$ is very small, i.e.~close
to the light-cone, where the most singular term in the light-cone operator
product expansion of the interacting light-cone Hamiltonian densities
will contribute.  The divergence is regulated by introducing a spatial UV
cutoff
$a^2$, where $ma \ll 1$,  and letting $G( m^2 x^2 )\rightarrow
G(m^2 (x^2 + a^2))$.  Then the most singular part of eqn.~\etwo, comes from the
exponential $\exp \beta^2 G(x)\approx \exp [-ln m^2(x^2 +a^2)/4\pi]$.
The singular term is isolated by restricting the $d^2 x$ integral to the
region $x^2 < l^2$, where $ml \ll 1$ and $l > a$,
\eqn\sing{
-i{\alpha^2\over 2\beta^4}\int^0_{-\infty} dx^+
\int_{x^2 < l^2}  dx^- \left[ m^2 (x^2 + a^2)
\right]^{-\beta^2 \over 4\pi}\ .
}
This expression differs from the equal-time result\samuel\mrg\ in that
the space-time integral is over regions with $x^+ < 0$, verses regions
with $t < 0$ in the equal-time case.  One might therefore
worry that there might be a cancellation between
space-like and time-like regions for the light-cone field theory
case. The space-like region for $x^+ < 0 $ is
$x^- > 0$, and it can be parametrized as $x^- = -r e^{+\theta}$ and
$x^+ = -r e^{-\theta}$.  The measure is just $\int_{0}^{l/\sqrt 2} rdr
\int_{-\infty}^{\infty}d\theta$.  Similarly for the time-like
region $x^- <0$, the parametrization is $x^- = -re^{+\theta}$ and
$x^+ = -r e^{-\theta}$, and the measure is the exactly the same as the
space-like region.  The Green's functions are independent of
$\theta$, and the space-like and time-like contributions add.  The light-cone
result
is the same as the equal-time result; isolating the $a$ dependence of
eqn.~\sing,
\eqn\asing{
\cE^-_2 \sim -i{\alpha^{2(1- {\beta^2\over4\pi})}\over \beta^4}
\int d\theta
{ a^{2(1-{\beta^2\over 4\pi} )} \over (1-\beta^2 / 4\pi )} \ .
}
This result is valid for all $\beta^2 \neq 4\pi$.  As $a\rightarrow 0$, the
expression vanishes for $\beta^2 < 4\pi$, and has a power-law divergence for
$\beta^2 >4\pi$.
At $\beta^2 = 4\pi$, the $a$ dependence is really $\ln a$, i.e.~$\cE^-_2$
diverges logarithmically as $a\rightarrow 0$.
Note that this logarithmic divergence does {\it not} occur at the phase
transition point $\beta^2 = 8\pi$.  And according to
ref.~\mrg, the $2p$ point contribution in the equal time-case diverges
logarithmically at $\beta^2 /8\pi = 1- (2p)^{-1}$ and the odd $p$-point
functions are UV finite for all $\beta^2 < 8\pi$.  For $\beta^2 < 8\pi$,
the divergences are less severe as $p$ increases; for $\beta^2 \geq
8\pi$ they become worse, and in this sense perturbation theory breaks down.

It is clear from the
above analysis that vacuum bubbles in light-cone field theory,
which correspond to the $k^+ = 0$ region of momentum space, are
in general non-vanishing.  And one proper way of evaluating them is to use
the coordinate space diagram approach, which follows directly from the
Dyson expansion in perturbation theory.

\newsec{The $k^+ = 0$ contribution to the Feynman propagator}

The vacuum energy density is not a physical observable for the sine-Gordon
model (no gravity) so its singularity properties are a relatively
mild
concern.
What is more important are singularities in physical
connected Green's functions, which from the equal-time
analysis\mrg\agg, occur only at $\beta^2 = 8\pi$.
To study this we will calculate to second order the in light-cone perturbation
theory the connected two-point Green's function (the Feynman propagator)
$\Gamma^{(2)} (k)$,
\eqn\Feynman{
\Gamma^{(2)} (k) = \int d^2 x e^{i k\cdot x}
\lstate 0 T \left[  \phi (x) \phi (0) \exp {\left( -i\int dx^+ P^-_{\rm int}
\right) } \right] \, {\state 0}_{\rm conn} \ .
}
The leading order contribution $\Gamma^{(2)}_0$ is $\int d^2 x e^{ik\cdot x}
G(x)= G_0 (k)$,
\eqn\gkay{
\en{
G_0 (k) =& {i\Theta(k^+ )
\over 2k^+ \left[ k^- -m^2/(2k^+ ) +i\epsilon \right] }
+ {i\Theta(- k^+ )
\over 2k^+ \left[ k^- -m^2/(2k^+ ) -i\epsilon \right] } \cr
=& {i\over \left[ k^+ k^- - m^2 / 2 + i \epsilon \right] } \ . \cr}
}
The next non-vanishing term is order $\alpha^2$.  The simplest
way to calculate it is to temporarily exponentiate $\phi (x)\phi (0)$
to $\exp{(a\phi (x))}\exp{(b\phi (x))}$, use eqn.~\wick, and then
pick out the order $ab$ terms;
\eqn\gammatwo{
\en{
\Gamma^{(2)}_2 (k)= G^2_0 (k) {4\alpha^2\over \beta^4}
\int d^2 x \biggl\{& \left[ \cosh {(\beta^2 G(x))}
-\beta^4 G^2(x) -1 \right] \cr
&- e^{-ik\cdot x} \left[ \sinh {(\beta^2 G(x))}
-\beta^2 G(x) \right] \biggr\} \ .\cr}
}
This equation is manifestly covariant and equivalent to the equal-time
result\mrg.

\topfigure {sgfig2.eps} {2} {The $x^+$ ordered light-cone diagrams for the
order $\alpha^2$ connected Green's function $\Gamma^{(2)}_2$
that correspond to the $[\cosh \cdots ]$ term in eqn.~\gammatwo.}

Now we can interpret it in terms of light-cone diagrams.  After integrating
over $x^-$, the $[\cosh \cdots ]$ term of eqn.~\gammatwo\ is given by
\eqn\cbubbles{
\en{
G^2_0 (k) & {\alpha^2\over 2\beta^4}
\sum_{n=2}^{\infty} {\beta^{2n} \over (2n)!} \int^{\infty}_0
\prod_{p=1}^{n} {dk^+_p \over k^+_p} \delta (\sum_{q=1}^{n} k^+_q )
\cr &
\int^{\infty}_{-\infty} dx^+ \left\{
\Theta ( -x^+ )
e^{ i x^+ \sum_{r=1}^{n} {m^2\over 2 k^+_r} }
+
\Theta ( x^+ )
e^{ -i x^+ \sum_{r=1}^{n} {m^2\over 2 k^+_r} } \right\}
\ .}
}
These are ``connected bubble" diagrams as shown in figure 2a-b.
Integrating over $x^+$ will generate $1/(P^-_0 - i\epsilon )$
denominators.
Like the
true vacuum bubble diagrams of eqn. \twolc\ and fig.~1, they receive
support only in the $k^+ = 0$ region of light-cone momentum.
Similarly, the  $[\sinh \cdots ]$ term of eqn.~\gammatwo\ has the
light-cone momentum space expansion
\eqn\zdiagrams{
\en{
G^2_0 & (k) {\alpha^2\over 2\beta^4}
\sum_{n=1}^{\infty} {\beta^{2n+1} \over (2n+1)!} \int^{\infty}_0
\prod_{p=1}^{n} {dk^+_p \over k^+_p} \int^{\infty}_{-\infty} dx^+ \Biggl\{
\Theta ( - x^+ ) \delta (k^+ - \sum_{q=1}^{n} k^+_q )
\cr &
e^{ - i x^+ \left(  k^- - \sum_{r=1}^{n} {m^2\over 2 k^+_r} \right) } +
\Theta ( x^+ ) \delta (k^+ + \sum_{q=1}^{n} k^+_q )
e^{  - i x^+ \left( k^- + \sum_{r=1}^{n} {m^2\over 2 k^+_r} \right) }
\Biggr\} \ .}
}
The $\Theta ( x^+ ) \cdots$ term of this expression are the
``Z"-diagrams of fig.~3b.
They vanish for all physical (i.e.~particles moving forward in time)
momenta $k^+ > 0$, because of the constraint $\delta (k^+ + \sum_p k^+_p )$.
The regular light-cone diagrams of fig.~3a contain a contribution
from the $k^+ = 0$ region of each internal momentum, since like the
bubble diagrams, their denominators are singular when
$\prod_p k^+_p ( \sum_r m^2 / 2 k^+_r - k^- )$ vanishes.  The singularity
occurs when two internal light-cone momenta simultaneously vanish while the
constraint $\sum_q k^+_q = k^+$ is still preserved.  (Note that this
is impossible for $\phi^3$ theory\changma. )
Therefore both the connected bubble diagrams of fig.~2a-b and the regular
light-cone diagrams of fig.~3a have contributions from the $k^+ = 0$
region.  These contributions do not cancel between the two types of diagrams,
because they occur to different orders in $\beta$ perturbation
theory.  The connected bubble diagrams are even order in $\beta$, while the
regular diagrams are odd order.

\topfigure {sgfig3.eps} {3} {The $x^+$ ordered light-cone diagrams for the
order $\alpha^2$ connected Green's function $\Gamma^{(2)}_2$
that correspond to the $[\sinh \cdots ]$ term in eqn.~\gammatwo.
In particular, the ``Z"-diagrams of  fig.~3b vanish
for physical ($k^+ > 0$) external light-cone momentum.}

We can verify the assertion that the connected bubble diagrams of fig.~2a-b
contribute to the connected Green's function $\Gamma^{(2)}_2$
by considering the non-perturbative UV divergence of
the sine-Gordon model that arises from integrating over the small $|x|$ region
of eqn.~\gammatwo.
Separately, the sum of connected bubble diagrams of fig.~2 and the
sum of the diagrams of fig.~3a diverge at $\beta^2 = 4\pi$.  This
is easy to see; the sources of the divergence are the $e^{\beta^2 G (x)}$
terms in eqn.~\gammatwo, and the result follows from the
analysis for the true bubble diagrams of eqn.~\twolc.  However, when the
terms are combined and the net divergence of eqn.~\gammatwo\ is
considered, the result for the regulated contribution
from the small $|x|$ region is
\eqn\newsing{
\Gamma^{(2)}_2 (k)= G^2_0 (k)
 \sim {\alpha^{2(1- {\beta^2\over4\pi})}\over \beta^4}
|x|^2 \int d\theta
{ a^{2(1-{\beta^2\over 8\pi} )} \over (1-\beta^2 / 8\pi)}\ .
}
This is valid for all $\beta^2 \neq 8\pi$; for $\beta^2 = 8\pi$ the
net contribution diverges logarithmically as $a\rightarrow 0$.
The important point is that without the $k^+ = 0$ contribution
from the connected bubble diagrams one gets the false, and unphysical,
result that the phase transition is at $\beta^2 = 4\pi$.
The interplay between regular light-cone diagrams and the
connected bubble diagrams that produces the correct non-perturbative (in
$\beta$)
UV divergence of the sine-Gordon model is expected to occur order by order in
$\alpha$ perturbation theory\agg.

\newsec{Discussion}

The light-cone vacuum of the sine-Gordon model is unstable at
$\beta^2 \geq 8\pi$, by a simple application of Coleman's variational
technique to light-cone field theory.  The crucial element in the
light-cone analysis is that the small and large $k^+$ cutoffs required
for the variational calculation are related by parity.  This introduces
free-field mass dependence into the regulated expressions of
the vacuum expectation values of light-cone energy and momentum.

This momentum cutoff prescription is not a sufficient regulator
of light-cone perturbation theory;
the $k^+ = 0$ region of light-cone perturbation theory contributes
to connected Green's functions in sine-Gordon model perturbation theory,
and incorrect results will occur if this region of momentum space is
discarded.
This region contributes
to connected diagrams at the two-loop order. (The one loop contributions
are tadpoles, which are eliminated by normal ordering.)  The
existence of the four and higher point vertices in the theory is crucial to
have non-vanishing connected bubble diagrams and $k^+ = 0$ singularities
in the regular connected diagrams; $\phi^3$ theory
does not have any contribution from the
$k^+ = 0$ region to connected diagrams\changma.

While in this paper, we focused on the effect of $k^+ = 0$
digrams on the propagator,  all of the vertices that occur in the sine-Gordon
model also receive contributions from $k^+ = 0$ type diagrams.
Therefore, it is clear that the naive process of implementing a sharp cutoff
of the $k^+ = 0$ region without making any corrections to the interaction
light-cone Hamiltonian in perturbation theory
is doomed to failure.  With the sharp cutoff, corrections to the propagator
and vertices will have to be made order by order in perturbation theory;
neglecting the $k^+ = 0$ region will in general violate gauge or Lorentz
symmetries, and
the counterterms are necessary to restore them.  It is not clear,
however, that symmetry restoration provides strong enough constraints to
recover a unique
Hamiltonian.  Therefore, the light-cone Hamiltonian with cutoff
of the $k^+ = 0$ region is a phenomenological construction that needs to be
tuned order by order in perturbation theory.

Symmetry restoration in ``naive" light-cone perturbation theory has already
been
considered\blone\bltwo\ by Burkardt and Langnau.  In particular, in
ref.~\bltwo, the non-covariant two loop term that is added to light-cone
perturbation theory is a $k^+ = 0$ contribution.  In their analysis, they
recovered this term by starting with the covariant
momentum space approach and correctly integrating over $k^-$.
I have argued here that light-cone field theory
is completely valid at the level of the Dyson perturbative expansion, and that
neglecting $k^+ =0$ diagrams when integrating over $x^\pm$ is a source
of similar problems.

One may ask how these results can be applied to more physical models.
Gauge theories quantized on the light-cone in light-cone gauge $A_- = 0$
have an effective four point Fermi coupling which arises from
the constraint equation for $A_+$\ksop.  Because of its non-local
nature, the vertex carries an extra $1/ (k^+ )^2$.
Therefore for QED as well as QCD, a potential contribution from the
$k^+ = 0$ region via the connected bubble type diagrams exists starting
at two loop order.

We have noted that the coordinate space expression for the light-cone
vacuum energy density that comes directly from the Dyson perturbative
expansion is perfectly well defined diagram by diagram, while the
``equivalent" momentum space expression suffers from $k^+ = 0$ singularities.
The mathematical source of this disparity is the integral over $x^-$ that
generates the constraint $\delta (\sum k^+ )$.  It might be worthwhile
to try to regulate this distribution, and therefore alter the light-cone
vertices, so as to make bubble diagrams well defined in momentum space
light-cone field theory.

In principle, it might also be possible to build a sharp $k^+$ cutoff into the
theory at the Langrangian level.  This is the approach suggested in
ref.~\mrobertson\ in the context of discrete light cone quantization\bpa.
It would be interesting to study how this approach can recover
the $k^+ = 0$ region of regular light-cone perturbation theory.

\ack{
This analysis was partially motivated by discussions with M.~Burkardt
on Lorentz non-covariant counterterms in light-cone perturbation theory.
I would also like to thank D.~Robertson for explaining the results of
ref.~\mrobertson\ to me,
and C.~Thorn for many useful comments regarding light-cone field theory.
}

\noindent
Note added:\
K.~Hornbostel has brought to my attention
ref.~\hv, in which the variational technique of sec.~2  is applied to $\phi^4$
theory in two dimensions.  Their analysis uses a
light-cone momentum regularization scheme which, in the limit of
infinite cutoff $\Lambda$, reduces to the parity symmetric prescription
advocated here.

\vfill\eject
\listrefs
\bye